\documentclass[journal]{IEEEtran}
\usepackage{cite}
\usepackage{amsmath,amssymb}
\usepackage{amsfonts}
\usepackage{algorithmic}
\usepackage{graphicx}
\usepackage{textcomp}
\usepackage{physics}
\usepackage{bm}
\usepackage{multirow}
\usepackage[T1]{fontenc}
\usepackage{enumerate}
\usepackage{booktabs}
\usepackage{balance}
\usepackage{url}
\usepackage{threeparttable}
\usepackage[super]{nth}
\makeatletter
\let\MYcaption\@makecaption
\makeatother
\usepackage[font=footnotesize]{subcaption}

\makeatletter
\let\@makecaption\MYcaption
\makeatother
\usepackage{caption}

\def\BibTeX{{\rm B\kern-.05em{\sc i\kern-.025em b}\kern-.08em
    T\kern-.1667em\lower.7ex\hbox{E}\kern-.125emX}}

\begin{document}
\title{Radar-based Measurement of the Body Movements of Multiple Students in Classroom Environments}

\author{Yu~Oshima, Tianyi~Wang, Masaya~Kato, Haruto~Kobayashi, Itsuki~Iwata, Yuji~Tanaka, Shuqiong~Wu, Manabu~Wakuta, Masako~Myowa, Tomoko~Nishimura, Atsushi~Senju, and Takuya~Sakamoto
\thanks{Y.~Oshima, M.~Kato, H.~Kobayashi, I.~Iwata, and T.~Sakamoto are with the Department of Electrical Engineering, Graduate School of Engineering, Kyoto University, Kyoto 615-8510, Japan}
\thanks{T.~Wang is with the Institute for Multidisciplinary Sciences, Yokohama National University, 79-7 Tokiwadai, Hodogaya-ku, Yokohama 240-8501, Japan}
\thanks{Y.~Tanaka is with the Department of Electrical and Mechanical Engineering, Nagoya Institute of Technology, Nagoya 466-0061, Japan}
\thanks{S.~Wu is with the Department of Intelligent Media, 
The Institute of Scientific and Industrial Research, Osaka University, Osaka 567-0047, Japan}
\thanks{M.~Wakuta is with the Research Department, Institute of Child Developmental Science Research, Hamamatsu 430-0929, Japan}
\thanks{M.~Myowa is with the Division of Interdisciplinary Studies in Education, Graduate School of Education, Kyoto University, Kyoto 606-8504, Japan}
\thanks{T.~Nishimura and A.~Senju are with the Research Center for Child Mental Development, Hamamatsu University School of Medicine, Hamamatsu 431-3192, Japan}}

\maketitle

\begin{abstract}
We demonstrate the feasibility of the radar-based measurement of body movements in scenarios involving multiple students using a pair of 79-GHz millimeter-wave radar systems with array antennas. We quantify the body motion using the Doppler frequency calculated from radar echoes. The measurement accuracy is evaluated for two experimental scenarios, namely university students in an office and 
elementary school students in a classroom. The body movements measured using the two radar systems are compared to evaluate the repeatability and angle dependency of the measurement. Moreover, in the first scenario, we compare the radar-estimated body movement with subjective evaluation scores provided by two evaluators. In the first scenario, the coefficient of correlation between the radar-estimated body movement and the subjective evaluation score is 0.73 on average, with a maximum value of 0.97; in the second scenario, the average correlation coefficient of body movements measured using two radar systems is as high as 0.78. These results indicate that the proposed approach can be used to monitor the body movements of multiple students in realistic scenarios.
\end{abstract}

\IEEEpeerreviewmaketitle

\section{Introduction}
\label{sec:introduction}
Childhood and adolescence are critical stages of development, often marked by a higher prevalence of psychological and mental health challenges\cite{child1}. Detecting these issues relies heavily on self-report questionnaires, as objective measurement remains elusive. 
Among these challenges, emotional and behavioral problems frequently co-occur during childhood, posing unique difficulties for assessment. The Strengths and Difficulties Questionnaire (SDQ) has become a widely used tool to evaluate such issues, capturing emotional problems, hyperactivity/inattention, peer relationship difficulties, and more\cite{SDQ}. Notably, hyperactivity tendencies stand out as a dimension that lends itself particularly well to objective evaluation, offering a promising avenue for more precise assessment.

Therefore, objectively estimating students' body movements in the classroom could help identify hyperactivity tendencies that are required to assess the emotional and behavioral development status. While contact-type sensors are often favored to measure the body movements \cite{bm_contact1, bm_contact2, bm_contact3, bm_contact4}, they are not suitable for long-term use due to potential skin discomfort, irritation, and psychological distress. Non-contact methods, such as using video data \cite{video1, video2, video3} and audio recordings \cite{audio1, audio2} to measure body movements, have also been explored. However, these methods raise privacy concerns, even with multiple security measures in place, making them less practical for classroom use.

Unlike these conventional systems, radar-based technology \cite{radar1, radar2, radar3, radar4, radar5, radar6, resp1} does not require any sensors to be attached to a person's body, which helps prevent skin discomfort and irritation. Additionally, radar-based systems do not capture visual or audio data, thereby avoiding privacy concerns. Most existing research on radar-based human measurement has focused on minimizing body movements, as these movements can interfere with the accuracy of measuring respiration and heartbeats. Consequently, studies have aimed to reduce or eliminate the impact of body movements \cite{bm_remove1, bm_remove2, bm_remove3, bm_remove5, bm_remove6, bm_remove7, bm_remove8, bm_remove9, bm_remove10}. In summary, body movements are often viewed as a source of noise that decreases the effectiveness of radar-based measurements, leading to extensive research on methods to mitigate these effects.

By contrast, other studies have explored the radar-based measurement of body movements. Among them, many studies \cite{bm3, bm11, bm8, bm5, bm6, bm7, bm12, bm13, bm14, bm15, bm16, bm17, bm18} are focused on radar-based classification of motions and activities such as walking, running, and falling. For instance, Luo et al. \cite{bm3} proposed a machine-learning-based method to distinguish people walking from people running using radar data, where participants were instructed to behave in a specific way within a predefined scenario. Ding et al. \cite{bm11} developed a radar-based method for identifying types of motion, such as falling, foot stomping, jumping, squatting, walking, and jogging. Song et al. \cite{bm8} proposed a method of distinguishing 10 types of action and reconstructing the posture model from radar data.

In contrast, some studies are focused on the radar-based measurement of the intensity/frequency of the body movements \cite{bm10, bm2,bm9,bm1,bm4}. Ding et al. \cite{bm10} developed a radar-based method of detecting the posture and body movements associated with careless driving that can lead to traffic accidents. Lee et al. \cite{bm2} reported the measurement of healthy and attention-deficit/hyperactivity disorder participants using radar and accelerometer sensors, although the measurements were conducted in clinical settings only and not in realistic settings. As discussed above, there are not many studies reported on the radar-based measurement of the intensity/frequency of the body movements. Therefore, it remains an open question whether radar can be used to measure the body movements of multiple people in realistic scenarios, such as those of students in a classroom. 

In this study, we investigate the feasibility of radar-based measurement of the intensity/frequency of the body movements by conducting radar measurements in realistic scenarios to estimate the body movements of multiple individuals. Two experimental scenarios were prepared to demonstrate the applicability of the proposed method to a wide range of the population having different ages. In the first scenario, the body movements of university students were measured using radar systems while the students were instructed to watch a video. In this scenario, subjective evaluations based on items in the SDQ were conducted during the radar measurements, enabling a comparison between radar-measured body movements and subjective evaluations related to the body movements of each student. 

In the second scenario, the body movements of elementary school students were measured using radar systems during daily class activities in a classroom. Part of the data recorded in the second scenario has been reported in prior work \cite{bm4}, where range–angle radar echo intensity maps were generated to detect large-scale body movements larger than the range resolution of the radar system (i.e., 44.7 mm). By contrast, the present study focuses on small-scale body movements as small as a fraction of the wavelength (i.e., 3.8 mm), which means that even a tiny movement less than 0.1 mm could be detected in the present study. Another difference between the prior study \cite{bm4} and this study is that the latter demonstrates the simultaneous monitoring of body movements of six students, whereas the former demonstrated the monitoring of only two students. In addition, this study is the first to verify the consistency of the intensity of the body movements of multiple students measured using two radar systems located at different locations.

\section{Radar Measurement of Body Motion}\label{method}
In this study, we use a radar system with array antennas that are approximated by an $N$-element linear array with element spacing $\lambda/2$, where $\lambda$ is the wavelength. Let $s_n(t,r)$ denote the signal received by the $n$-th virtual element ($n=1,\cdots,N$) for slow time $t$ and range $r=\tau/2c$, where $\tau$ is the fast time and $c$ is the speed of light.

Through beamforming \cite{beamfomer}, we obtain a complex radar image 
\begin{align}
I^{\prime}_\mathrm{C}(t,r,\theta)=\sum_{n=1}^N \alpha_{n}w_{n}(\theta)s_{n}(t,r),
\end{align}
where $\theta$ is the azimuth angle, $\alpha_n$ is a Taylor window coefficient, and $w_n(\theta)=\mathrm{e}^{\mathrm{j} \pi n \sin \theta}$ is the beamformer weight.
Next, we define $L$ time segments of length $T$ as $\ell T \leq t \leq (\ell+1)T$ $(\ell=0,1,\cdots,L-1)$, where we set $T=30$ s. For the $\ell$th time segment, the static clutter is suppressed by subtracting the time-averaged complex radar image as 
\begin{align}
I_{\mathrm{C}}(t,r,\theta)=I^{\prime}_\mathrm{C}(t,r,\theta)-\frac{1}{T}\int_{\ell T}^{(\ell+1)T}I^{\prime}_\mathrm{C}(t',r,\theta)\mathrm{d}t'
\end{align}
for $\ell T \leq t \leq (\ell+1)T$.

Then, for the $\ell$th time segment, we estimate the target position $\left(r_{\mathrm{}}^{(\ell)},\theta_{\mathrm{}}^{(\ell)}\right)$ represented in polar coordinates as
\begin{align}
  \label{eq:Time_ave}
  \left(r_{\mathrm{}}^{(\ell)},\theta_{\mathrm{}}^{(\ell)}\right)=\arg\max_{(r,\theta)}I_{\ell}(r,\theta),
\end{align}
where the $\ell$th power radar image $I_{\ell}(r,\theta)$ is defined by
\begin{equation}
I_{\ell}(r,\theta)=\frac{1}{T}\int_{\ell T}^{(\ell+1)T}|I_\mathrm{C}(t,r,\theta)|^2\dd t,
\end{equation}
which means that the target position estimate is updated every 30 s.

The body displacement $d(t)$ of the target person is then estimated as
$d(t)=(\lambda/4\pi)\mathrm{unwrap}\{{\angle{I_{\mathrm{C}}(t,r_{\mathrm{max}}^{(\ell)},\theta_{\mathrm{max}}^{(\ell)})}}\}$,
where $\ell$ is selected depending on $t$ satisfying $\ell T \leq t \leq (\ell+1)T$, $\angle$ denotes the phase of a complex number, and $\mathrm{unwrap}\{{\cdot}\}$ represents an unwrapping operator that adds or subtracts integer multiples of $2\pi$ so that the difference between adjacent phase samples is less than $\pi$.

We propose a body movement index $b(t)$ to evaluate the intensity and frequency of the body movements as
\begin{align}
  b(t)=\sqrt{\frac{1}{T_\mathrm{b}}\int_{t-T_\mathrm{b}/2}^{t+T_\mathrm{b}/2}\left|\frac{\rm{d}}{\rm{d}\tau}d(\tau) \right|^2 \mathrm{d}\tau},
\end{align}
where $T_\mathrm{b}=0.5$ s represents the time length for evaluating the body movements. The proposed $b(t)$ corresponds to the root mean square of the Doppler velocity, which is proportional to the Doppler frequency. Few studies have reported the validation of the radar-measured body movements compared with subjective evaluation scores of a standard questionnaire. To our knowledge, this study is the first to investigate the feasibility of the radar-based measurement of body movements that might be linked to the psychological status of the target person.

\section{Experimental Evaluation of Body Movement Measurements}
We performed radar measurements and subjective evaluations of body movements of multiple people in two scenarios. The first and second scenarios were university and elementary school students, respectively, seated in a classroom. For the first scenario, the radar-measured body movement is compared with the subjective evaluation index recorded by questionnaire.

\subsection{Scenario 1: University Students}
\label{exp1}
\subsubsection{Radar Measurement}
\begin{figure}[bt]
    \centering
    \includegraphics[width=0.7\linewidth]{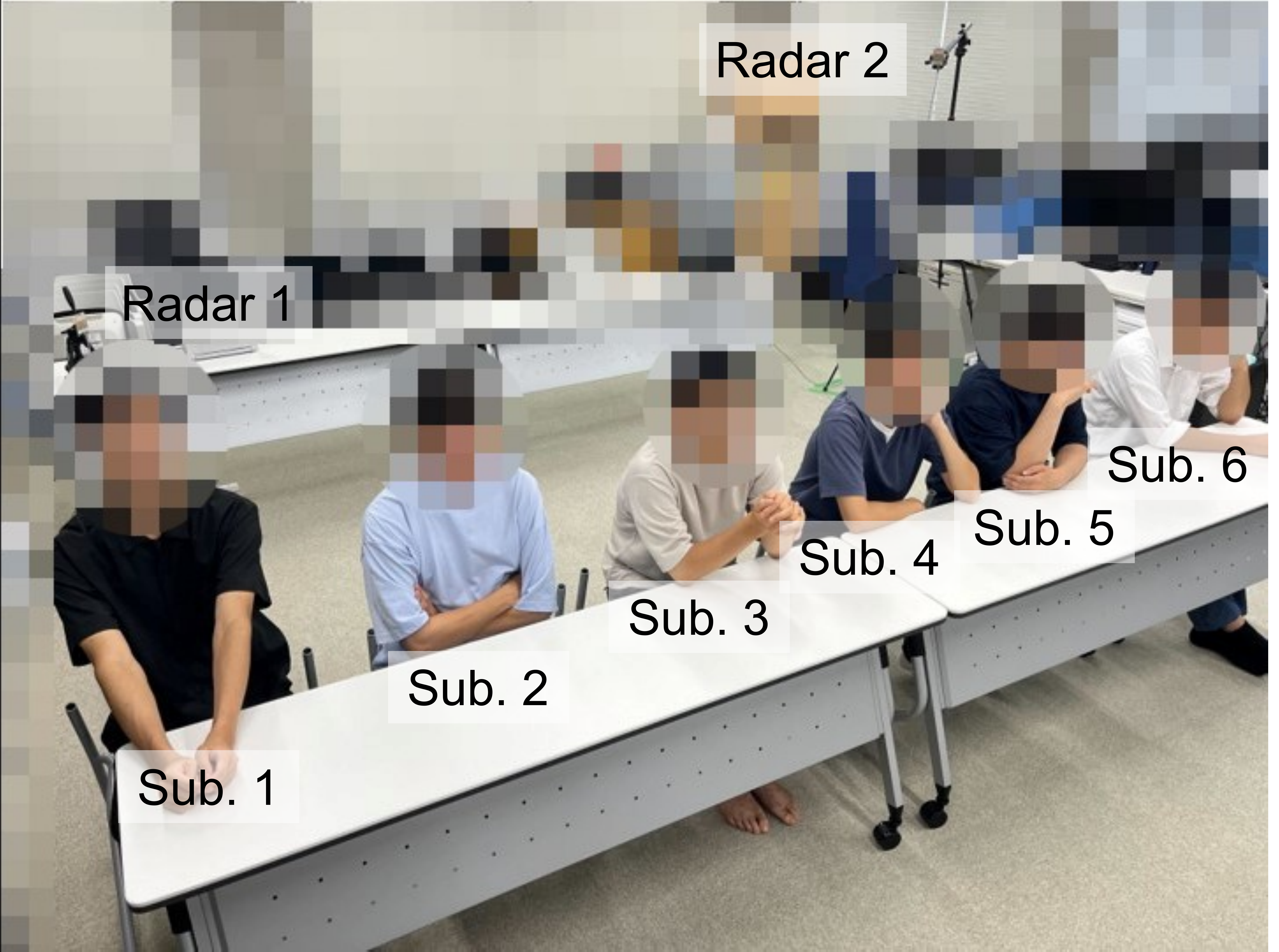}
  \caption{Experimental setup of scenario 1.}
  \label{fig:setup}
\end{figure}

\begin{figure}[bt]
    \centering
    \includegraphics[width=0.9\linewidth]{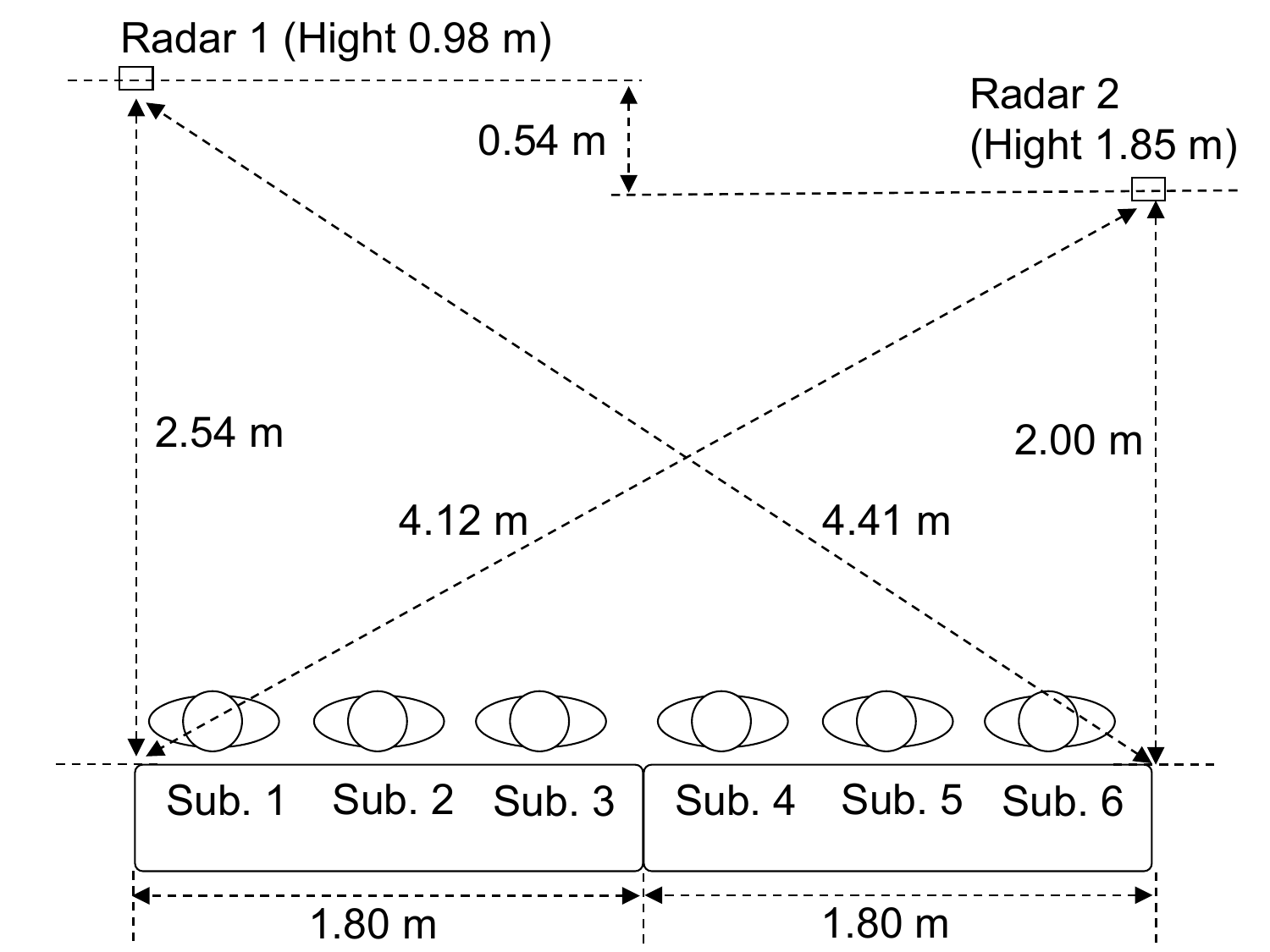}
  \caption{Schematic plan view of experimental setup of scenario 1.}
  \label{fig:setup2}
\end{figure}

We first performed measurements using $K=2$ radar systems for $M=6$ participants who are selected randomly from 13 people, who were healthy adult university students aged between 21 and 27 years. Radar systems 1 and 2 were placed to the left rear and right rear of the participants, respectively. Figure \ref{fig:setup} and \ref{fig:setup2} show the measurement setup and its schematic plan view with the radar systems positioned behind the participants. Each measurement lasted 15 minutes. We conducted measurements $J=9$ times with different sets of participants.  During the measurements, participants were instructed to remain seated and to watch a video shown on a large screen.

The radar specifications are given in Table \ref{tab:1}. A pair of frequency-modulated continuous-wave radar systems with a center frequency of 79.0 GHz and a bandwidth of 3.6 GHz were used. The radar system had three transmitting elements and four receiving elements forming a multiple-input and multiple-output array that can be approximated by a virtual linear array with 12 elements with an element spacing of $\lambda/2$. 

\begin{table}[bt]
  \begin{center}
      \caption{RADAR SYSTEM SPECIFICATIONS}
      \label{tab:1}
      \begin{tabular}{cc}
          \toprule
          Parameter                               & Value                     \\ \midrule
          Modulation                              & FMCW \\
          Center frequency                        & $79.0~\mathrm{GHz}$       \\
          Center wavelength                       & $3.8~\mathrm{mm}$         \\
          Bandwidth                               & $3.6~\mathrm{GHz}$        \\
          No. of transmitting (Tx) antennas       & $3$                       \\
          No. of receiving (Rx) antennas          & $4$                       \\
          Tx element spacing                      & $7.6~\mathrm{mm}$         \\
          Rx element spacing                      & $1.9~\mathrm{mm}$         \\
          Beamwidths of Tx elements (E-/H-planes) & $\pm 4^\circ/\pm33^\circ$ \\
          Beamwidths of Rx elements (E-/H-planes) & $\pm 4^\circ/\pm45^\circ$ \\
          Range resolution                        & $44.7~\mathrm{mm}$        \\
          Sampling frequency (slow time)          & $100~\mathrm{Hz}$         \\
          \bottomrule
      \end{tabular}
  \end{center}
\end{table}

\begin{figure}[bt]
      \centering
      \includegraphics[width=0.9\linewidth]{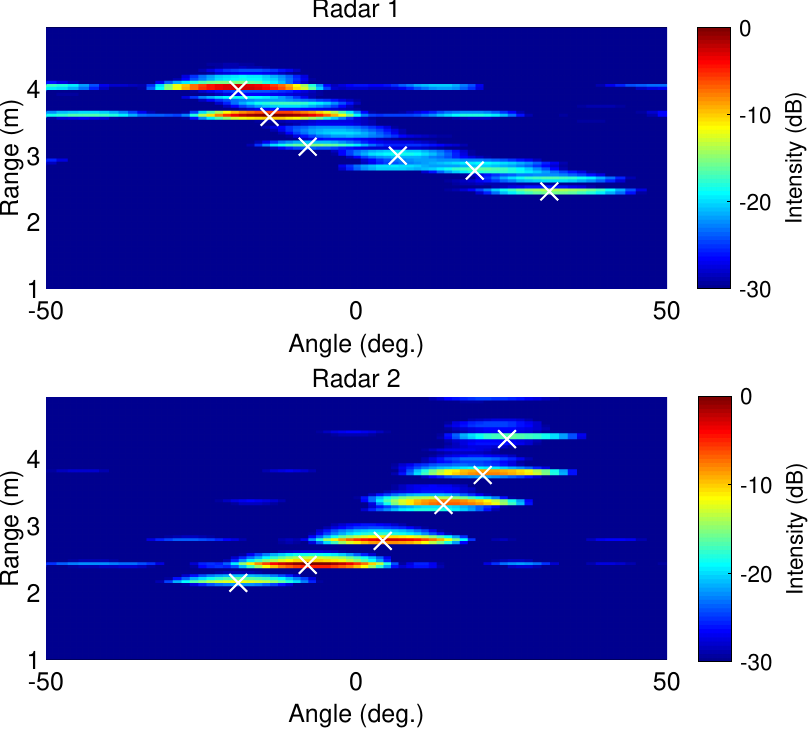}
      \caption{Example of radar images $I_{\ell}(r,\bm{\theta})$ for $\bm{\ell} = 29$ in experiment 1 obtained from radar systems 1 and 2 (scenario 1).}
      \label{fig:radarimage}
\end{figure}

Figure \ref{fig:radarimage} shows an example of a pair of radar images $I_{\ell}(r,\theta)$ for $\ell=29$ in experiment 1 obtained from the radar systems. In the upper and lower panels of the figure, we see six dominant peaks corresponding to the six participants in the experiment, where the angle and range for each participant differ for radar systems 1 and 2 because the positions and angles of the radar systems are not identical.

The position of the $m$th participant is estimated by finding the local maxima as
\begin{equation}
\left(r_m^{(\ell)},\theta_m^{(\ell)}\right) = \arg\max_{(r,\theta)\in R_m}I_{\ell}(r,\theta)
\end{equation}
for $m=1,2,\cdots,M$, where $R_m$ is a rectangular region that includes the $m$th participant's seating position. As described in Section II, the estimated position of each participant was updated every $T_\mathrm{a}=30$ s. The estimated positions of the six participants are shown by crosses in Fig. \ref{fig:radarimage}.

\begin{figure}[bt]
      \centering
      \includegraphics[width=0.9\linewidth]{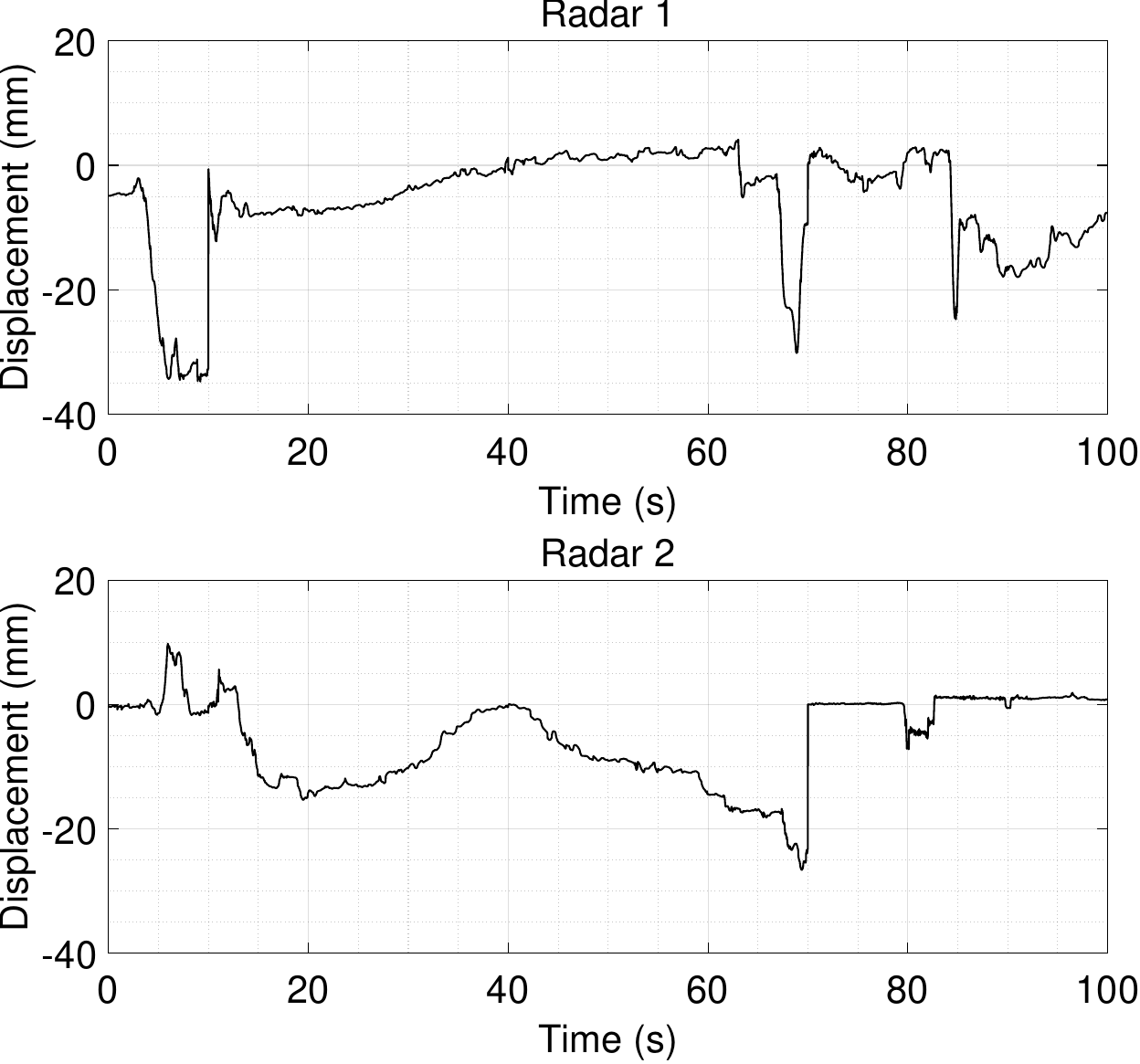}
      \caption{Example of body displacement estimated using radar systems 1 and 2 (scenario 1).}
      \label{fig:disp}
\end{figure}

The body displacement $d(t)$ estimated using radar systems 1 and 2 is shown in Fig. \ref{fig:disp}. The displacement estimates vary between the two radar systems owing to differences in the radar echoes, which depend on the part of the target body reflecting the radar signal according to the relative position and orientation of the target body and radar antenna. However, overall body movements are captured using both radar systems 1 and 2, indicating the possibility of the evaluation of intensity/frequency of body movements.

\begin{figure}[bt]
  \begin{center}
    \centering
    \includegraphics[width=0.99\linewidth]{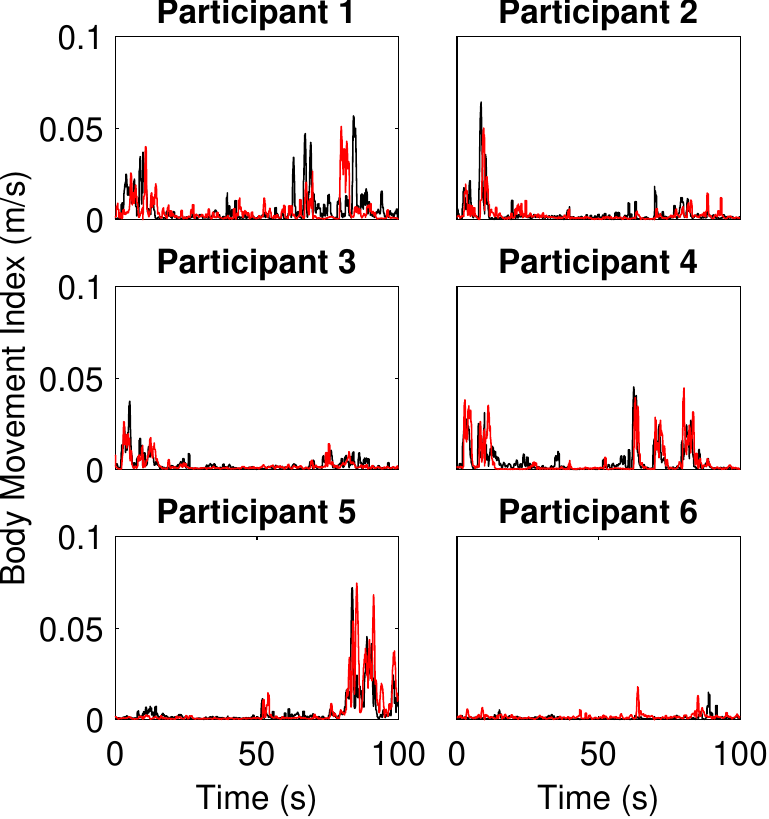}
    \caption{Body movement indices $b^{(1)}_m(t)$ (black line) and $b^{(2)}_m(t)$ (red line) measured using two radar systems in experiment 1 (scenario 1).}
    \label{fig:bm_result}
  \end{center}
\end{figure}

The body movement index $b(t)$ of the $m$th participant measured using the $j$th radar system is denoted by $b^{(j)}_m(t)$ $(m=1,2,\cdots,M; j=1,2)$. An example of this measurement is shown in Fig. \ref{fig:bm_result}, where the black and red lines represent the estimates $b^{(1)}_m$ and $b^{(2)}_m$ obtained from radar systems 1 and 2, respectively. We see that the body movements measured using radar systems 1 and 2 are in good agreement.  

Table \ref{tab:2} gives the correlation coefficient \begin{align}
    \label{eq:soukan}
    \rho_{m,m',\ell}=
    \frac{\int_{\ell T_\rho}^{(\ell+1)T_{\rho}} \hat{b}_{m}^{(1)}(t)\hat{b}_{m'}^{(2)}(t)\;\mathrm{d}t}{\sqrt{\int_{\ell T_\rho}^{(\ell+1)T_{\rho}} \left|\hat{b}_{m}^{(1)}(t)\right|^2\;\mathrm{d}t} \sqrt{\int_{\ell T_\rho}^{(\ell+1)T_{\rho}} \left|\hat{b}_{m'}^{(2)}(t)\right|^2\;\mathrm{d}t}}
\end{align}
for $\hat{b}^{(1)}_m = b^{(1)}_m-\overline{b}_{m}^{(1)}$ and $\hat{b}^{(2)}_{m'}=b^{(2)}_{m'}-\overline{b}_{m'}^{(2)}$ measured over the $\ell$th time segment $\ell T_\rho\leq t\leq (\ell+1)T_\rho$ $(\ell=0,\cdots,L-1)$. $\overline{b}_{m}^{(1)}$ and $\overline{b}_{m'}^{(2)}$ are the averages of $b_{m}^{(1)}(t)$ and $b_{m'}^{(2)}(t)$, respectively. In this study, we set $T_\rho=60$ s and $L=15$ because the measurements were performed over $LT_\rho=15$ minutes in each experiment. From Table \ref{tab:2}, we see that the average correlation of the radar-measured body movement indices obtained from radar systems 1 and 2 was 0.56. These results show that the body movements measured using two radar systems located at different positions are consistent to some extent although they are not completely identical. The difference of the body movement indices is partly due to the fact that different body parts show different movements, which means that the intensity and frequency of the motion of the right and left arms are reasonably correlated but not completely identical.

\begin{table}[bt]
  \centering
  \caption{Correlation coefficient $(1/L)\sum_{\ell=0}^{L-1}\rho_{m,m',\ell}$ of the body movement index measured using two radar systems (scenario 1)}
  \label{tab:2}
  \begin{tabular}{c|c|c|c|c|c|c|c}\hline
  \toprule
      \multicolumn{2}{c|}{} & \multicolumn{5}{c}{Radar 2}\\
      \cline{3-8}
      \multicolumn{2}{c|}{}& 1 & 2 & 3 & 4 & 5 & 6 \\ \hline
          \multirow{6}{*}{Radar 1}
          &1& 0.41 & 0.01 & 0.14 & $-0.06$ & 0.26 & 0.26\\ \cline{2-8}
          &2& 0.25 & 0.67 & 0.03 & 0.19 & $-0.03$ & $-0.06$\\ \cline{2-8}
          &3& 0.07 & $-0.05$ & 0.69 & 0.21 & 0.15 & 0.04\\ \cline{2-8}
          &4& $-0.16$ & 0.15 & 0.13 & 0.42 & $-0.07$ & 0.09\\ \cline{2-8}
          &5& $-0.05$ & 0.07 & 0.28 & 0.34 & 0.44 & 0.05\\ \cline{2-8}
          &6& 0.02 & $-0.03$ & 0.23 & 0.07 & 0.36 & 0.72\\ \hline
          \toprule
  \end{tabular}
\end{table}

When the movements of multiple people are measured using multiple radar systems simultaneously, it is important to associate echoes to improve the monitoring accuracy \cite{resp1}. Specifically in this study, for each echo detected by radar system 1, we identify a corresponding echo among those detected by radar system 2 using the correlation coefficient $\rho_{m,m',\ell}$. We evaluate the accuracy $p$ in associating echoes as 
\begin{equation}
p = \frac{1}{ML}\sum_{m=1}^{M}\sum_{\ell=0}^{L-1}\hat{p}_{m,\ell},
\end{equation}
where $\hat{p}_{m,\ell}$ is defined by
\begin{equation}
\hat{p}_{m,\ell} = \left\{
\begin{array}{ll}
1 &(\mathrm{if}\;\rho_{m,m,\ell} \geq \rho_{m,m',\ell}\; \mathrm{for}\;\forall m'\neq m)\\
0 &(\mathrm{otherwise}).
\end{array}\right.
\end{equation}
Table \ref{tab:3} gives the accuracy $p$ in each experiment. We see that the accuracy $p$ is higher than 63\% except for experiment 4, and the average accuracy is 68.3\%.

\begin{table}[bt]
  \begin{center}
      \caption{Accuracy $p$ of target association when using radar systems 1 and 2 (scenario 1)}
      \label{tab:3}
      \begin{tabular}{cccccc}
          \toprule
          Experiment no. & 1 & 2 & 3 & 4 & 5 \\ 
          Target assoc. accuracy $p$ (\%)& 67.8 & 83.3 & 75.6 & 36.7 & 82.2  \\
          \midrule
          Experiment no. & 6 & 7 & 8 & 9 & \\        
          Target assoc. accuracy $p$ (\%)& 65.6 & 71.1 & 63.3 & 68.9 & \\
          \bottomrule
      \end{tabular}
  \end{center}
\end{table}

\subsubsection{Subjective Evaluation}
The intensity and frequency of the body movements of each participant were evaluated subjectively by $K'=2$ evaluators during the measurements. The evaluators were requested to score the two items for each participant:
\begin{itemize}
\item $\bar{\beta}_1$: Restlessness, overactivity, inability to stay still for long.
\item $\bar{\beta}_2$: Constant fidgeting or squirming.
\end{itemize}
These items were taken from the SDQ \cite{SDQ} and respectively evaluated with scores $\bar{\beta}_1$ and $\bar{\beta}_2$, each on a scale from 0 to 2, where 0 indicates "not true", 1 indicates "somewhat true", and 2 indicates "certainly true." The summation $\bar{\beta}=\bar{\beta}_1+\bar{\beta}_2$ $(0\leq \bar{\beta}\leq 4)$ is used in the analysis below.

First, the score $\bar{\beta}$ is normalized to remove the evaluators' individual differences.
Let $\bar{\beta}^{(m,j,k)}$ denote the score $\bar{\beta}$ for the $m$th participant in the $j$th experiment given by the $k$th evaluator. The scores then normalized and averaged as 
\begin{align}
\displaystyle\beta^{(m,j)} = \frac{1}{K'}\sum_{k'=1}^{K'}\frac{\bar{\beta}^{(m,j,k')}}{\displaystyle\frac{1}{MJ}\sum_{m'=1}^M \sum_{j'=1}^J \bar{\beta}^{(m',j',k')}},
\end{align}
where $M=6$ is the number of participants, $J=9$ is the number of experiments, and $K'=2$ is the number of evaluators.

To compare with the subjective index of body movements $\beta^{(m,j)}$, let us define an objective radar-measured body movement index $b^{(m,j)}$ of the $m$th participant in the $j$th experiment by averaging the body movement indices obtained from the $K=2$ radar systems as
\begin{align}
    b^{(m,j)}=\frac{1}{K}\sum_{k=1}^{K}\sqrt{\frac{1}{LT_{\rho}}\int_{0}^{LT_{\rho}}\left|\hat{b}_{m}^{(j)}(t)\right|^2\mathrm{d}t},
\label{eq:bmj}
\end{align}
where the radar number $k\in\{1, 2\}$ is not explicitly shown on the right-hand side of (\ref{eq:bmj}).

\begin{figure}[bt]
    \centering
    \includegraphics[width=0.9\linewidth]{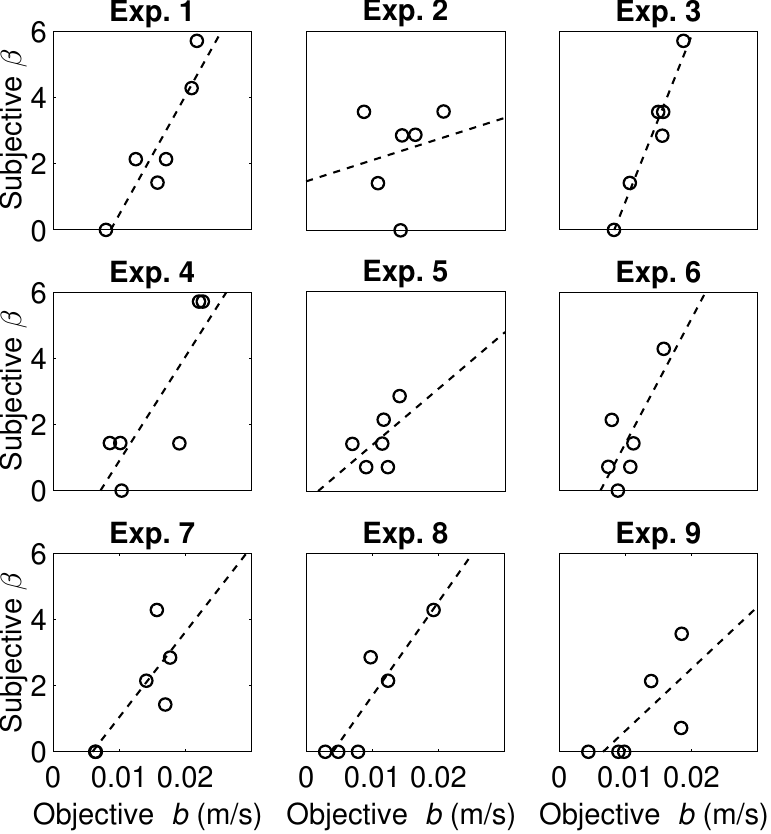}
    \caption{Scatter plot of the radar-measured objective body movement index $b$ and subjective evaluation score $\bm{\beta}$ (scenario 1).}
    \label{fig:sca2}
\end{figure}

Figure \ref{fig:sca2} is a scatter plot of $b^{(m,j)}$ and $\beta^{(m,j)}$ for each experiment. The correlation coefficients for these scatter plots are summarized in Table \ref{tab:4},
The correlation coefficient of $b^{(m,j)}$ and $\beta^{(m,j)}$ was 0.73 on average. In experiment 3, in particular, the correlation coefficient was as high as 0.97. In contrast, in experiment 2, the correlation coefficient was as low as 0.19.
This is partly because one of the participants was almost stationary without any movements, but made large movements a few times, which captured the human evaluators' attention and lead erroneous high score of the subjective body movement index. When excluding this exceptional case in experiment 2, the correlation coefficient was 0.80 on average. We confirm from these results that the subjective evaluation scores given by the evaluators can be predicted to some extent using the proposed radar-measured body movement index.

\begin{table}[bt]
{\begin{center}
      \caption{Correlation coefficient of the radar-measured body movement index $b$ and subjective evaluation score $\bm{\beta}$ (scenario 1)}
      \label{tab:4}
      \begin{tabular}{cccccc}
          \toprule
          Experiment number $j$ & 1 & 2 & 3 & 4 & 5 \\ 
          Correlation coeff. of $b$ and $\beta$ & 0.91 & 0.19 & 0.97 & 0.83 & 0.52  \\
          \midrule
          Experiment number $j$ & 6 & 7 & 8 & 9 & \\ 
          Correlation coeff. of $b$ and $\beta$ &  0.76 & 0.79 & 0.91 & 0.72 & \\   \bottomrule
      \end{tabular}
  \end{center}
          }
\end{table}

As described above, the body movements of multiple people were measured using radar systems in a noncontact manner, and the repeatability of measurements was evaluated by comparing data from the two radar systems. The radar-measured body movement index was then compared with the subjective evaluation made by human evaluators with a standard questionnaire, demonstrating the correlation between the objective radar measurement and subject evaluation scores. 

\subsection{Scenario 2: Elementary School Students}
\label{exp2}
\begin{figure}[bt]
      \centering
      \includegraphics[width=0.7\linewidth]{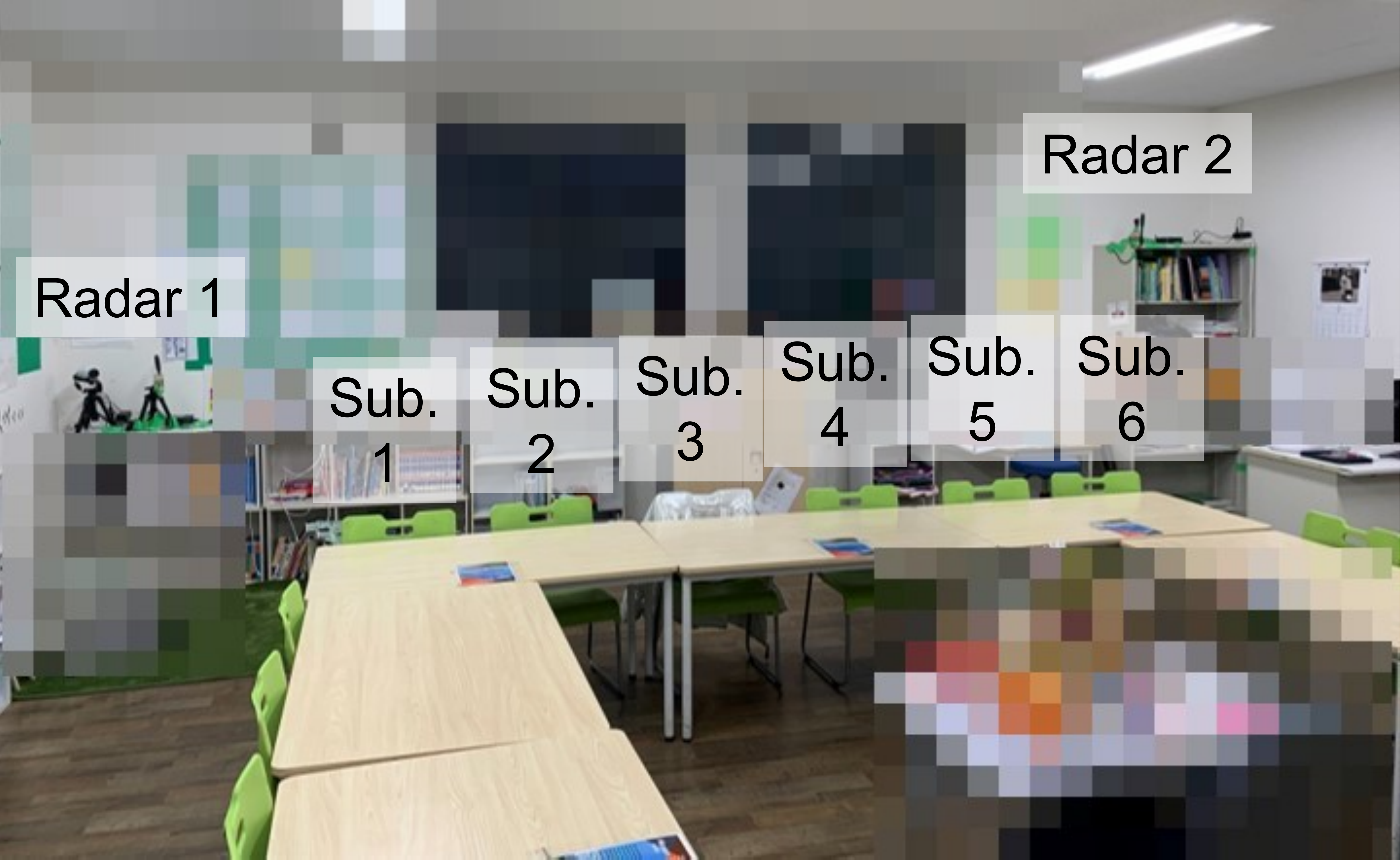}
      \caption{Experimental setup of scenario 2.}
      \label{fig:setup_school}
\end{figure}

In scenario 2, we used the same radar systems as used in scenario 1 for the measurement of body movements in an elementary school classroom over 9 days. Students were 8 or 9 years old and carried out their regular classroom activities without any restrictions. We evaluated the body movement index of $M=6$ students seated in a U-shaped seating pattern as shown in Fig. \ref{fig:setup_school}. The time spent on activities in the U-shaped seating arrangement varied by day, and therefore, the time length used for radar signal analysis differed each day, as summarized in Table \ref{tab:5}.

\begin{table}[bt]
  \begin{center}
      \caption{Total measurement time in the U-shaped seating arrangement (scenario 2)}
      \label{tab:5}
      \begin{tabular}{cccccc}
          \toprule
          Day & 1 & 2 & 3 & 4 & 5 \\ 
          Measurement time (min) & 62 & 94 & 137 & 42 & 91  \\
          \midrule
          Day & 6 & 7 & 8 & 9 & \\        
          Measurement time (min) & 35 & 23 & 43 & 71 & \\
          \bottomrule
      \end{tabular}
  \end{center}
\end{table}

As shown in Fig. \ref{fig:setup_school}, two radar systems were installed at the back of the classroom to verify the repeatability of the radar-measured body movement index $b(t)$. 
Radar system 1 was positioned in the left corner of the classroom, positioned at a height of 0.90 m, and located 5.10 m from the left end and 2.50 m from the right end of the table. 
Radar system 2 was installed in the right corner of the classroom, set at an angle to look down on the children from a height of 1.78 m, and located 3.60 m from the left end and 4.70 m from the right end of the table. Moreover, at the conclusion of the 9-day experiment, two teachers filled in the SDQ for each participant. 
Our previous study \cite{bm4} already reported the relationship between the body movements measured using radar systems and the SDQ scores, and we therefore avoid reiterating the discussion in this paper.

\begin{figure}[bt]
  \begin{center}
    \centering
    \includegraphics[width=0.99\linewidth]{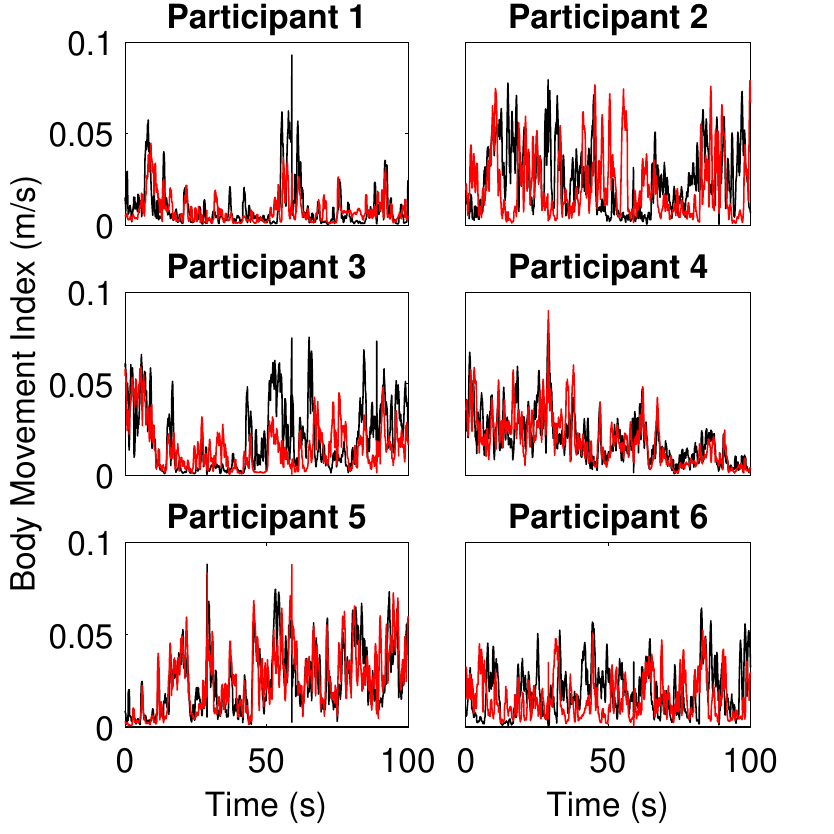}
    \caption{Body movement indices $b^{(1)}_m(t)$ (black line) and $b^{(2)}_m(t)$ (red line) measured using two radar systems in experiment 1 (scenario 2).}
    \label{fig:bm_result_school}
  \end{center}
\end{figure}

The body movement index $b(t)$ measured using the radar systems is shown in Fig. \ref{fig:bm_result_school}. The black and red lines represent the estimates obtained from radar systems 1 and 2, respectively. Except for a few participants, the body movement indices measured using radar systems 1 and 2 show good agreement. Note that different movements are captured by the two radar systems if the participants make asymmetric body movements because the systems are installed apart from each other. A comparison of Figs. \ref{fig:bm_result} and \ref{fig:bm_result_school} shows that elementary school students make more intense and frequent body movements than the university students. 

Next, as in Section \ref{exp1}, the average correlation coefficient $(1/L)\sum_{\ell=0}^{L-1}\rho_{m,m,\ell}$ of the body movement indices $b(t)$ obtained from the two radar systems for the $m$th participant is calculated (Table \ref{tab:6}). The high average correlation coefficient of 0.78 indicates the repeatability of the two radar systems. Table \ref{tab:7} shows that the average accuracy in associating echoes detected by the two radar systems is as high as 82.2\%. Note that the accuracy value for day 8 is missing because there were no continuous 15-minute data for the U-shaped seating arrangement on that day.

\begin{table}[bt]
  \begin{center}
      \caption{Average correlation coefficients $(1/L)\sum_{\ell=0}^{L-1}\rho_{m,m,\ell}$ of the body movement index for the $m$th participant obtained using two radar systems (scenario 2)}
      \label{tab:6}
      \begin{tabular}{ccccccc}
          \toprule
          Participant no. & 1 & 2 & 3 & 4 & 5 & 6  \\ \midrule
          Correlation coeff. & 0.68 & 0.48 & 0.83 & 0.89 & 0.93 & 0.84  \\
          \bottomrule
      \end{tabular}
  \end{center}
\end{table}

\begin{table}[bt]
  \begin{center}
      \caption{Target association accuracy $p$ when using radar systems 1 and 2 (scenario 2)}
      \label{tab:7}
      \begin{tabular}{cccccc}
          \toprule
          Day & 1 & 2 & 3 & 4 & 5 \\ 
          Target assoc. accuracy $p$ (\%)& 60.0 & 98.9 & 82.2 & 74.4 & 73.3  \\
          \midrule
          Day & 6 & 7 & 8 & 9 & \\        
          Target assoc. accuracy $p$ (\%)& 90.0 & 97.8 & - & 81.1 & \\
          \bottomrule
      \end{tabular}
  \end{center}
\end{table}

\section{Discussion}
The results shown above suggest that radar-based systems have the potential to collaborate, argument, and possibly replace conventional methods of
evaluating students' hyperactivity, which are currently performed by teachers and counselors. This could pave the way for automatic and objective assessments of emotional and behavioral development.
However, several challenges need to be addressed to achieve this goal:
\begin{itemize}
\item \textbf{Angular Dependency}: The body movement indices from the two radar systems showed a correlation coefficient of 0.78, indicating that the data were not perfectly aligned (with 1.00 being ideal). This implies that body movement detection varies based on the angle of observation. For example, when a student’s right hand moves while the left hand remains still, the radar system on the right side captures the movement more effectively than the radar system on the left side.
\item \textbf{Identifying Individuals}: The current method cannot distinguish between students, which can lead to incorrect associations of body movement data when students change seating positions. For this purpose, radar-based individual identification techniques can be applied.
\item \textbf{Comprehensive Assessment}: The SDQ used for evaluating students uses multiple aspects beyond body movements to evaluate hyperactivity. These additional factors require evaluation scores given by human evaluators such as school teachers and psychologists, which need to be integrated to the radar-based system to ensure a multi-faceted assessment.
\item \textbf{Environment Diversity}: This study carried out validation of the proposed method in two different scenarios; one involved university students in an office environment, while the other focused on elementary school students in a single classroom. Although analyses were conducted with six participants in both scenarios, it is still uncertain how factors such as classroom size and the number of students affect the consistency between subjective and objective evaluations or the consistency between the two radar systems. For this reason, further validation in more diverse classroom environments will be needed in the future.
\end{itemize}
By developing solutions to these challenges, the proposed radar-based method could enable automatic and objective evaluation of emotional and behavioral development status in the future.

\section{Conclusion}
In this study, we demonstrated the feasibility of radar-based body movement measurement in scenarios involving multiple students. The proposed body movement index is based on the root mean square of Doppler velocity obtained from radar echoes that are reflected off multiple target people. To verify the accuracy of the proposed method, we performed experiments in two distinct scenarios with two radar systems and two evaluators for the objective and subjective assessments of each participant's body movements. In the first experiment, body movement was estimated while having university students watch the same video, whereas in the second experiment, body movement was estimated over 9 days during elementary school classes. 
As a result, in the first experiment, the correlation coefficient of radar-estimated body movement and subjective evaluation scores was 0.73 on average, with a maximum value of 0.97. If one case is excluded as an exception, the average correlation coefficient was 0.80. Additionally, the accuracy in associating echoes measured using two radar systems was 68.3\% on average. In the second experiment, the average correlation coefficient of body movements measured using the two radar systems was 0.78, and the accuracy in associating echoes obtained using two radar systems was as high as 82.2\% on average. These results demonstrate that the proposed method is effective in monitoring the body movements of multiple students in realistic scenarios. 

\section*{Acknowledgment}
This work was supported in part by SECOM Science and Technology Foundation, in part by JST under Grants JPMJMI22J2 and JPMJMS2296, in part by JSPS KAKENHI under Grants 21H03427, 23H01420, 23K19119, and 24K17286, and in part by the New Energy and Industrial Technology Development Organization (NEDO). 
This work involved human subjects in its research. Approval of all ethical and experimental procedures and protocols was granted by the Ethics Committee of the Graduate School of Engineering, Kyoto University, under Approval No. 202219 and No. 202223.
Written informed consent for participation was obtained from the participants and their legal guardians, and all participants agreed to the privacy policy. 
We thank Ms. Hiroko Takagi (Kansai International Academy), Ms. Marwa Elgezery, MEd (Kansai International Academy), and Ms. Yoko Morisaki (Kansai International Academy) for their cooperation in our elementary school experiment and Glenn Pennycook, MSc, from Edanz (https://jp.edanz.com/ac) for editing a draft of this manuscript. 

\balance

\begin{IEEEbiography}[{\includegraphics[width=1in,height=1.25in,clip,keepaspectratio]{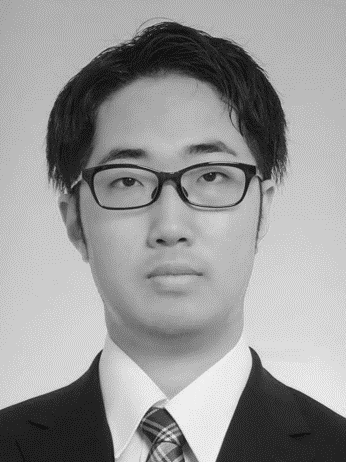}}]
{Yu Oshima} received a B.E. degree in electrical and electronic engineering from Kyoto University, Kyoto, Japan, in 2023. He is currently pursuing an M.E. degree in electrical engineering at the Graduate School of Engineering, Kyoto University. Mr. Oshima was a recipient of the Electronics Society Student Award in 2024.
\end{IEEEbiography}

\begin{IEEEbiography}[{\includegraphics[width=1in,height=1.25in,clip,keepaspectratio]{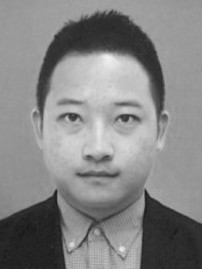}}]{Tianyi Wang} received his Ph.D. degree in Health Science from Osaka University, Japan, in 2020. He was a senior research at the Department of Robotics, Faculty of Science and Engineering, Ritsumeikan University, Japan from 2020 to 2022. He was a postdoctoral researcher at the Faculty of Frontier Engineering, Kanazawa University, Japan from 2022 to 2023. From 2023 to 2024, he was a program-specific researcher at the Department of Electrical Engineering, Graduate School of Engineering, Kyoto University. He is presently a specially appointed assistant professor at the Institute of Multidisciplinary Science, Yokohama National University. His research interests include engineering in healthcare science, human--robot interactions, human posture analysis, and the application of artificial intelligence for healthcare robots. He is affiliated with academic societies such as the IEEE Life Science Community, IEEE Young Professionals, Society for Nursing Science and Engineering, and Japan Society of Maternal Health.
\end{IEEEbiography}

\begin{IEEEbiography}[{\includegraphics[width=1in,height=1.25in,clip,keepaspectratio]{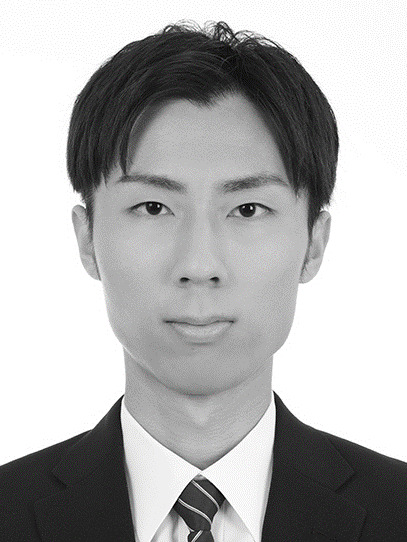}}]{Masaya Kato} received a B.E. degree in electrical and electronic engineering from Kyoto University, Kyoto, Japan, in 2023. He is currently pursuing an M.E. degree in electrical engineering at the Graduate School of Engineering, Kyoto University.
\end{IEEEbiography}

\begin{IEEEbiography}[{\includegraphics[width=1in,height=1.25in,clip,keepaspectratio]{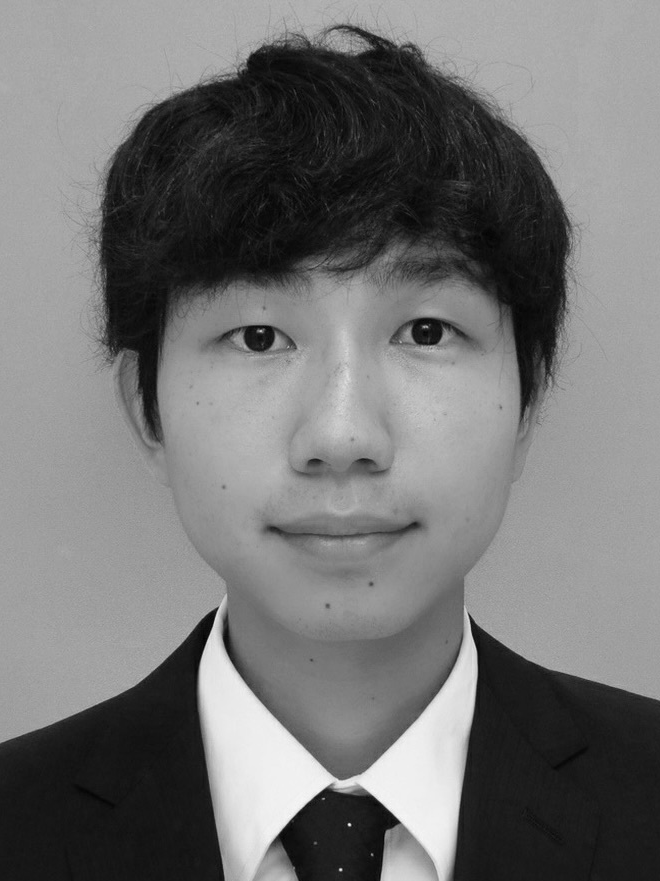}}]{Haruto Kobayashi}(Graduate Student Member, IEEE) received a B.E. degree in electrical and electronic engineering from Kyoto University, Kyoto, Japan, in 2023. He is currently pursuing an M.E. degree in electrical engineering at the Graduate School of Engineering, Kyoto University.
\end{IEEEbiography}

\begin{IEEEbiography}[{\includegraphics[width=1in,height=1.25in,clip,keepaspectratio]{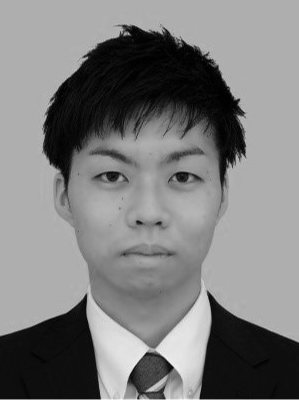}}]{Itsuki Iwata} received a B.E. degree in electrical and electronic engineering from Kyoto University in 2022 and an M.E. degree in electrical engineering from the Graduate School of Engineering, Kyoto University, in 2024. 
Mr. Iwata was a recipient of the IEEE Antennas and Propagation Society Kansai Joint Chapter Young Engineer Technical Meeting Best Presentation Award in 2021.
\end{IEEEbiography}

\begin{IEEEbiography}[{\includegraphics[width=1in,height=1.25in,clip,keepaspectratio]{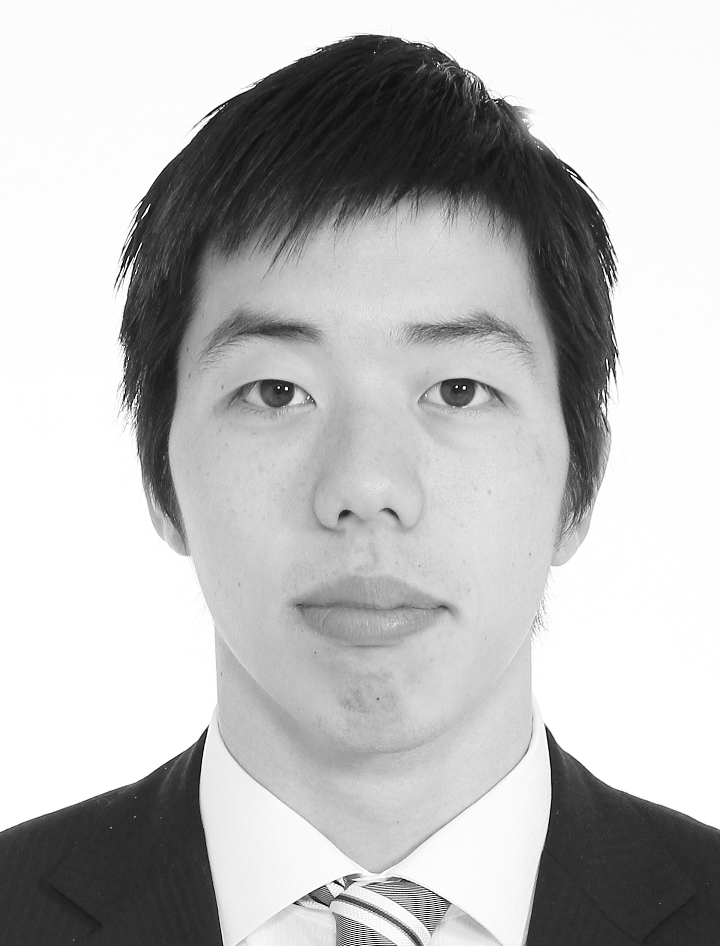}}]{Yuji~Tanaka} (Member, IEEE) received B.E., M.E., and Ph.D. degrees in engineering from Kanazawa University, Ishikawa, Japan, in 2015, 2017, and 2023, respectively. From 2017 to 2020, he worked at the Information Technology R\&D Center, Mitsubishi Electric Corporation. From 2023 to 2024, he worked as an assistant professor at the Graduate School of Engineering, Kyoto University. He is currently an assistant professor at the Graduate School of Engineering, Nagoya Institute of Technology. His research interests include radar signal processing, radio science, and the radar measurement of physiological signals. He was awarded Second Prize at the URSI-JRSM 2022 Student Paper Competition and the Young Researcher's Award from the Institute of Electrical and Electronics Engineers (IEICE) Technical Committee on Electronics Simulation Technology. He is a member of the IEICE and the Society of Geomagnetism and Earth, Planetary and Space Sciences (SGEPSS).
\end{IEEEbiography}

\begin{IEEEbiography}[{\includegraphics[width=1in,height=1.25in,clip,keepaspectratio]{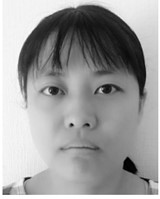}}]{Shuqiong~Wu} was born in Shanxi, China, in 1985. She received B.E. and M.E. degrees from BeiHang University, Beijing, China, in 2008 and 2011 respectively. She received her Ph.D. degree, majoring in computational intelligence and systems science, from the Tokyo Institute of Technology, Tokyo, Japan, in 2015. From 2015 to 2020, she was a research fellow with the Graduate School of Informatics, Kyoto University. Since 2020, she has been working as an assistant professor with SANKEN (The Institute of Scientific and Industrial Research), Osaka University. Her current research topics include dual-task-based cognitive impairment detection, cognitive status monitoring, medical image reconstruction, and contactless biometric sensing. Her research interests include biomedical signal processing, image processing, three-dimensional reconstruction, pattern recognition, and machine learning. 
\end{IEEEbiography}

\begin{IEEEbiography}[{\includegraphics[width=1in,height=1.25in,clip,keepaspectratio]{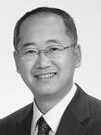}}]{Manabu~Wakuta} received a Ph.D. degree from Osaka University, Japan, in 2012 after working as a special education teacher over 20 years. He then became a chief researcher at the Institute of Child Developmental Science Research. Moreover, he is a visiting faculty member at Osaka University and a joint visiting researcher at the Hamamatsu University School of Medicine.
\end{IEEEbiography}

\begin{IEEEbiography}[{\includegraphics[width=1in,height=1.25in,clip,keepaspectratio]{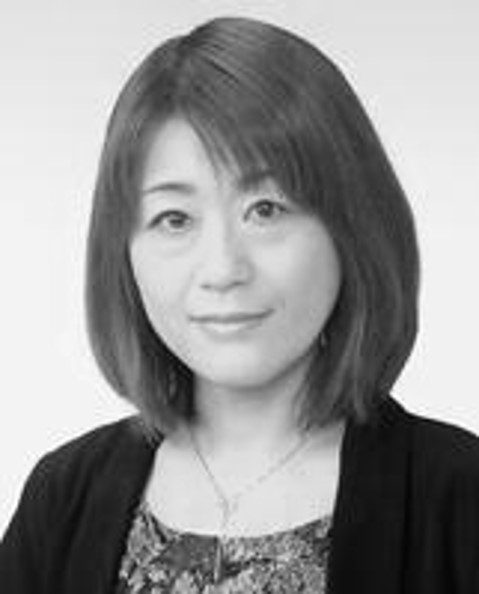}}]{Masako~Myowa} received her Ph.D. from Kyoto University, Japan, in 1999. From 2007 to 2014, she was an associate professor at Kyoto University. Since 2014, she has been a professor at Kyoto University. Her research interests include the emergence, development, and evolutionary foundations of human intelligence. She is the author of more than 200 publications, editorials, and books in the fields of primatology and developmental science. She received the Nakayama Encouragement Award, Takashima Award, and Award for International Contributions to Psychology, among others.
\end{IEEEbiography}

\begin{IEEEbiography}[{\includegraphics[width=1in,height=1.25in,clip,keepaspectratio]{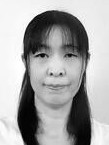}}]{Tomoko~Nishimura} received a Ph.D. degree from Osaka University, Japan, in 2016. She then worked at Osaka University as a project research fellow and moved to Hamamatsu University School of Medicine as a project assistant professor. Currently, she is a project lecturer at the Research Center for Child Mental Development, Hamamatsu University School of Medicine. Moreover, she is a research fellow at the Institute of Child Developmental Science Research.
\end{IEEEbiography}

\begin{IEEEbiography}[{\includegraphics[width=1in,height=1.25in,clip,keepaspectratio]{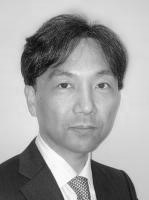}}]{Atsushi~Senju} received a Ph.D. degree from the University of Tokyo, Japan, in 2005. He then held various research positions at Birkbeck, University of London, including Leverhulme Trust Visiting Fellow, ESRC/MRC Postdoctoral Fellow, ESRC Research Fellow, MRC Career Development Award Fellow, Wellcome Trust/ISSF Research Fellow, and Reader in Social Neuroscience. He transitioned to his current position, Professor and Center Director at the Research Center for Child Mental Development, Hamamatsu University School of Medicine, in 2021. He was awarded the JSPS Prize (2020), Nakayama Award for Distinguished Early and Middle Career Contributions (2015), BPS Margaret Donaldson Early Career Prize (2015), and BPS Neil O’Connor Award (2011) and nominated as a Rising Star by the Association for Psychological Sciences (2011).
\end{IEEEbiography}

\begin{IEEEbiography}[{\includegraphics[width=1in,height=1.25in,clip,keepaspectratio]{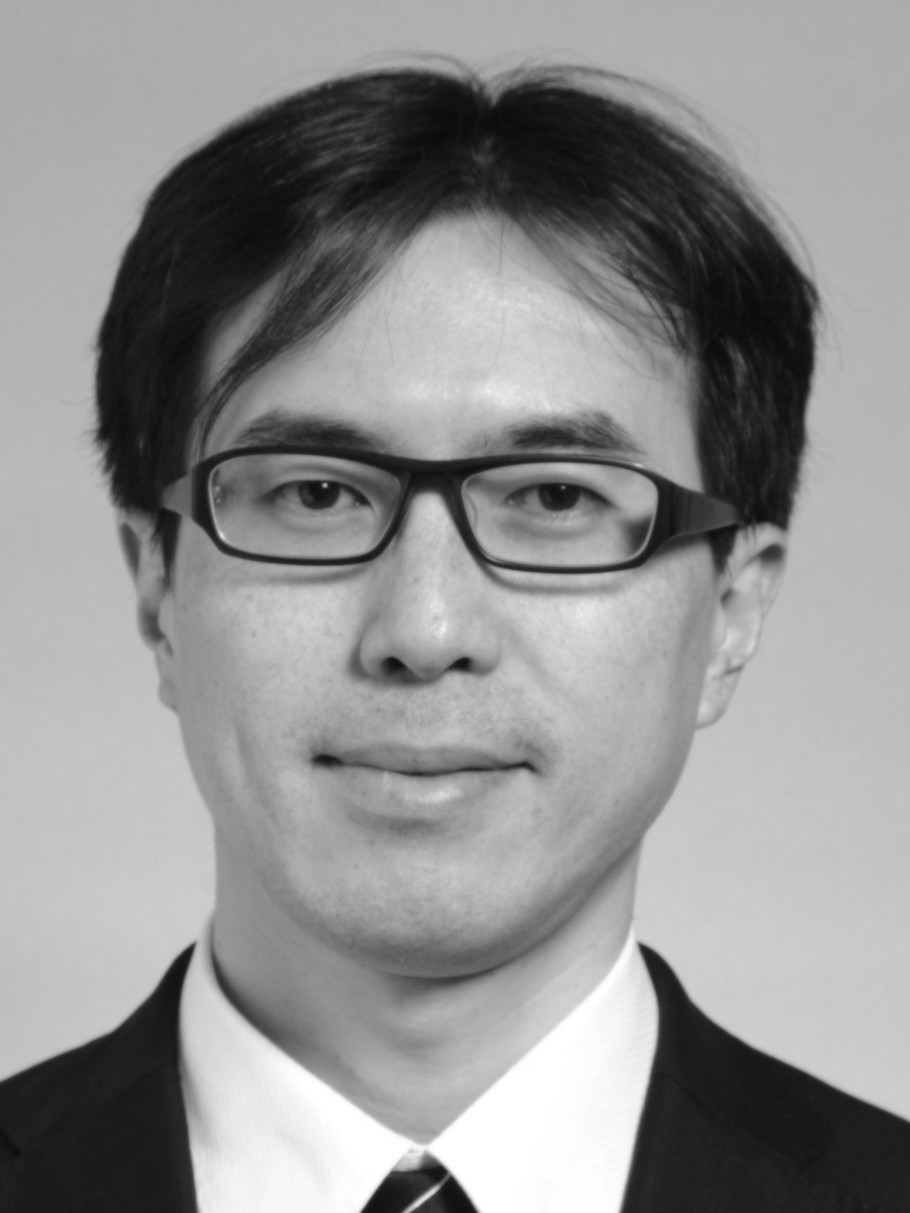}}]
    {Takuya Sakamoto} (Senior Member, IEEE) received a B.E. degree in electrical and electronic engineering from Kyoto University, Kyoto, Japan, in 2000 and M.I. and Ph.D. degrees in communications and computer engineering from the Graduate School of Informatics, Kyoto University, in 2002 and 2005, respectively. From 2006 through 2015, he was an assistant professor at the Graduate School of Informatics, Kyoto University. From 2011 through 2013, he was also a visiting researcher at Delft University of Technology, Delft, the Netherlands. From 2015 until 2019, he was an associate professor at the Graduate School of Engineering, University of Hyogo, Himeji, Japan. In 2017, he was also a visiting scholar at the University of Hawaii at Manoa, Honolulu, HI, United States. From 2019 until 2022, he was an associate professor at the Graduate School of Engineering, Kyoto University. From 2018 through 2022, he was a PRESTO researcher of the Japan Science and Technology Agency, Japan. Since 2022, he has been a professor at the Graduate School of Engineering, Kyoto University. His current research interests lie in wireless human sensing, radar signal processing, and the radar measurement of physiological signals.

   Professor Sakamoto was a recipient of the Best Paper Award from the International Symposium on Antennas and Propagation (ISAP) in 2004, Young Researcher's Award from the Institute of Electronics, Information and Communication Engineers of Japan (IEICE) in 2007, Best Presentation Award from the Institute of Electrical Engineers of Japan in 2007, Best Paper Award from the ISAP in 2012, Achievement Award from the IEICE Communications Society in 2015, 2018, and 2023, Achievement Award from the IEICE Electronics Society in 2019, Masao Horiba Award in 2016, Best Presentation Award from the IEICE Technical Committee on Electronics Simulation Technology in 2022, Telecom System Technology Award from the Telecommunications Advancement Foundation in 2022, and Best Paper Award from the IEICE Communication Society in 2007 and 2023.
\end{IEEEbiography}


\end{document}